\documentclass[sigconf]{acmart}
\usepackage[utf8]{inputenc}

\PassOptionsToPackage{table,xcdraw}{xcolor} 

\usepackage{textgreek}
\usepackage{multirow}
\usepackage{subcaption}
\usepackage{xcolor,colortbl}
\definecolor{gray}{rgb}{0.1,0.1,0.1}
\usepackage[T1]{fontenc}
\usepackage{graphicx}
\usepackage{tabularx}
\usepackage{longtable}
\usepackage{enumitem}

\AtBeginDocument{%
  \providecommand\BibTeX{{%
    \normalfont B\kern-0.5em{\scshape i\kern-0.25em b}\kern-0.8em\TeX}}}

\acmConference[Woodstock '18]{Woodstock '18: ACM Symposium on Neural
  Gaze Detection}{June 03--05, 2018}{Woodstock, NY}
\acmBooktitle{Woodstock '18: ACM Symposium on Neural Gaze Detection, June 03--05, 2018, Woodstock, NY}
\acmPrice{15.00}
\acmISBN{978-1-4503-XXXX-X/18/06}

\acmSubmissionID{123-A56-BU3}

\usepackage{xspace}     % Used for abbreviation spacing
\usepackage{xpunctuate} % Used for abbreviation spacing % New (approved) packages
\usepackage{xargs}      % Used for new commands with optional arguments
\usepackage{soul}       % Used for custom comments
\usepackage{color}      % Used for custom colors in comments % New packages for custom comments
\definecolor{LIGHTPINK}{RGB}{237,157,202}
\definecolor{LIGHTRED}{RGB}{210,121,121}
\definecolor{LIGHTORANGE}{RGB}{230,170,50}
\definecolor{LIGHTGOLD}{RGB}{210,194,121}
\definecolor{LIGHTGREEN}{RGB}{121,210,121}
\definecolor{LIGHTAQUA}{RGB}{121,206,210}
\definecolor{LIGHTBLUE}{RGB}{121,124,210}
\definecolor{LIGHTPURPLE}{RGB}{153,102,255}
\definecolor{RED}{RGB}{178,34,34}
\definecolor{GRAY}{RGB}{166,166,166}
\definecolor{WHITE}{RGB}{255,255,255}

\newcommandx{\jane}[2][1=] 
    {\setulcolor{LIGHTGREEN}{\ul{#1}} \textcolor{LIGHTGREEN}
    {[\textbf{Jane:} #2]}}
\newcommandx{\guest}[3][1=]
    {\setulcolor{LIGHTORANGE}{\ul{#1}} \textcolor{LIGHTORANGE} 
    {[\textbf{#2:} #3]}}

\begin{document}
\title{Design for Hope: Cultivating Deliberate Hope in the Face of Complex Societal Challenges}

\author{JaeWon Kim}
\affiliation{%
  \institution{University of Washington}
  \city{Seattle}
  \country{USA}}
\orcid{0000-0003-4302-3221}
\email{jaewonk@uw.edu}

\author{Jiaying "Lizzy" Liu}
\authornote{All authors contributed equally to this research.}
\affiliation{%
  \institution{University of Texas at Austin}
  \city{Austin}
  \country{USA}}
\orcid{0000-0002-5398-1485}
\email{jiayingliu@utexas.edu}

\author{Lindsay Popowski}
\authornotemark[1]
\affiliation{%
  \institution{Stanford University}
  \city{}
  \country{USA}}
\orcid{}
\email{popowski@stanford.edu}

\author{Cassidy Pyle}
\authornotemark[1]
\affiliation{%
  \institution{University of Michigan}
  \city{Ann Arbor, MI}
  \country{USA}}
\orcid{0000-0003-4578-0226}
\email{cpyle@umich.edu}

\author{Ahmer Arif}
\affiliation{%
  \institution{University of Texas at Austin}
  \city{}
  \country{USA}}
\orcid{0000-0003-2412-9776}
\email{ahmer@utexas.edu}

\author{Gillian R. Hayes}
\affiliation{%
  \institution{University of California, Irvine}
  \city{}
  \country{USA}}
\orcid{}
\email{}

\author{Alexis Hiniker}
\affiliation{%
  \institution{University of Washington}
  \city{}
  \country{USA}}
\orcid{}
\email{alexisr@uw.edu}

\author{Wendy Ju}
\affiliation{%
  \institution{Cornell University}
  \city{New York City}
  \country{USA}}
\orcid{0000-0002-3119-611X}
\email{wendyju@cornell.edu}

\author{Florian `Floyd' Mueller}
\affiliation{%
  \institution{Monash University}
  \city{Melbourne}
  \country{Australia}}
\orcid{0000-0001-6472-3476}
\email{floyd@exertiongameslab.org}

\author{Hua Shen}
\affiliation{%
  \institution{University of Washington}
  \city{Seattle}
  \country{USA}}
\orcid{}
\email{huashen@uw.edu}

\author{Sowmya Somanath}
% \authornotemark[1]
\affiliation{%
  \institution{University of Victoria}
  \city{Victoria}
  \country{Canada}}
\orcid{0009-0005-8580-5215}
\email{sowmyasomanath@uvic.ca}

\author{Casey Fiesler}
\affiliation{%
  \institution{University of Colorado Boulder}
  \city{}
  \country{USA}}
\orcid{}
\email{casey.fiesler@colorado.edu}

\author{Yasmine Kotturi}
\affiliation{%
  \institution{University of Maryland, Baltimore County}
  \city{Baltimore}
  \country{USA}}
\orcid{}
\email{kotturi@umbc.edu}

\renewcommand{\shortauthors}{JaeWon Kim, et al.}

\begin{abstract}
Design has the potential to cultivate hope in the face of complex societal challenges. These challenges are often addressed through efforts aimed at harm reduction and prevention---essential but sometimes limiting approaches that can unintentionally narrow our collective sense of what is possible. This one-day, in-person workshop builds on the first Positech Workshop at CSCW 2024 by offering practical ways to move beyond reactive problem-solving toward building capacity for proactive goal setting and generating pathways forward. We explore how collaborative and reflective design methodologies can help research communities navigate uncertainty, expand possibilities, and foster meaningful change. By connecting design thinking with hope theory, which frames hope as the interplay of ``goal-directed,'' ``pathways,'' and ``agentic'' thinking, we will examine how researchers might chart new directions in the face of complexity and constraint. Through hands-on activities including problem reframing, building a shared taxonomy of design methods that align with hope theory, and reflecting on what it means to sustain hopeful research trajectories, participants will develop strategies to embed a deliberately hopeful approach into their research.
\end{abstract}




\maketitle

\section{Introduction}

Design---through its tools, processes, and cognitive practices such as ~\cite{cross2006designerly}---can help people navigate moments when challenges feel insoluble or the future uncertain~\cite{schon2017reflective, Bottrell-2007-ResistanceResilienceSchooling-z, Seelig-2013-ReframingProblemInnovation-d}. As both a discipline and a way of thinking, design supports the generation and development of ideas that may lead to new and sometimes transformative directions~\cite{Zimmerman2007-ur}. In particular, design thinking creates space for reframing problems, helping us move beyond surface-level solutions and toward reimagining what might be possible~\cite{rylander2009design}. At its core, design draws on empathy and creativity~\cite{Bennett2019-mr}, qualities that encourage a sense of possibility and agency. In this way, the act of designing can offer more than just solutions to problems; it can support researchers, collaborators, and communities in working through uncertainty and sustaining efforts toward meaningful change~\cite{light2014structuring}.

This framing is especially relevant for researchers tackling the layered, often messy, societal challenges. Consider social media, for example: social media once exemplified technology's power to expand human capabilities, such as enabling remote connections and beyond. However, it is now often viewed as irreparably broken, with rising concerns about misinformation~\cite{suarez2021prevalence, wang2019systematic}, youth mental health~\cite{nesi2020impact, liu2024exploring}, online abuse~\cite{blackwell2017classification, mandryk2023combating}, and more~\cite{cinelli2021echo, tucker2018social}. For researchers trying to imagine something better, it is easy to feel stuck, as though the best we can do is reactively respond and mitigate symptoms, not expand possibilities. However, design research offers a different perspective: rather than optimizing existing platforms, it invites us to step back and ask, is the current paradigm of social media that few platforms such as Instagram and TikTok offer truly the best embodiments of social media's potential? What else could social media be? Does social media interface have to be two-dimensional? Could it take an entirely new form? What if we let go of default assumptions and generate alternatives from the ground up~\cite{kim2025socialmediafeellike, Blythe-2016-Anti-solutionistStrategiesFiction-p}? In this sense, design becomes especially relevant to CSCW researchers who share the core goal of reframing complex problems. 

But ``design for hope'' is \textit{not} about blind optimism or naive trust. Both design and hope~\cite{Snyder-2002-HopeTheory-i} require the disciplined and intentional work of finding the right direction: by questioning assumptions, setting ambitious but reachable goals, iterating thoughtfully, and continuously reassessing whether the direction remains aligned with those goals~\cite{schon2017reflective}. This process moves beyond blue-sky speculation toward generating concrete, achievable next steps. In this process, designers (broadly defined) engage in techniques that help expand the problem space and avoid fixation~\cite{Jansson-1991-DesignFixation-e}. They engage with communities and stakeholders, using empathy, participatory practices~\cite{spinuzzi2005methodology, sanders2002user, Liang-2024-170Co-designingYouth-h, kotturi2024peerdea}, and speculative exploration~\cite{AugerAuger-2013-SpeculativeDesignSpeculation-x, Zolyomi-2024-EmotionTranslatorDyads-d, Dunne-2024-SpeculativeEverythingDreaming-d,klassen2024black} not just to imagine what could be built, but to make those ideas actionable and contextually grounded. Further, approaches like participatory~\cite{spinuzzi2005methodology, sanders2002user, Liang-2024-170Co-designingYouth-h, kotturi2024peerdea}, speculative~\cite{AugerAuger-2013-SpeculativeDesignSpeculation-x, Zolyomi-2024-EmotionTranslatorDyads-d, Dunne-2024-SpeculativeEverythingDreaming-d, klassen2024black}, and critical design~\cite{bardzell2013critical}, make this work explicit, offering strategies for exploring not only what might work but also what might go wrong~\cite{shilton2018values, do2023s, klassen2022run}. They also resonate with frameworks like \textit{critical hope}~\cite{Grain-2017-SocialJusticeDespair-y, Cahill-2010-''dreaming''Hope-j}, which call for deep engagement with uncertainty and discomfort as part of working toward meaningful change.

This workshop aims to bring together researchers invested in confronting complex societal challenges through collaborative reflection and methodological innovation. Through hands-on activities such as problem reframing, taxonomy development, and community-engaged exercises, participants will explore how design can serve as both a mindset and a method for cultivating hope. The goal is not to solve problems in one shot but to open up new pathways for imagining and iterating toward better futures, together. In doing so, this workshop frames design not only as a tool for innovation and breakthroughs but also as a practice for sustaining agency and resilience amongst researchers and communities in the face of uncertainty and despair.

\subsection{Background and Motivation}
Hope---defined as ``\textit{the perception that one can reach desired goals}''~\cite{Snyder-2002-HopeTheory-i}---is central to human flourishing. It fuels resilience\cite{Senger-2023-HopesRelationshipPandemic-x}, enhances well-being~\cite{To2023-id}, and allows us to move forward even in the face of hardship~\cite{Belen-2020-InfluenceFearPathways-o}. The psychology of hope outlines two core processes: \textbf{\textit{pathways thinking}} (developing multiple routes to a goal) and \textbf{\textit{agentic thinking}} (the motivation to pursue those routes)~\cite{Snyder-2002-HopeTheory-i}. \textit{\textbf{Goal-setting}} sits at the center of both. Specifically, goals that are most conducive to hopeful processes are neither trivial nor impossible; they strike a balance between challenge and attainability. Notably, hope tends to reinforce itself: individuals who engage in hopeful thinking and processes are more likely to generate alternative pathways and recover more quickly when they encounter obstacles~\cite{Snyder-2002-HopeTheory-i}.

Design thinking naturally aligns with this structure of hope.  Generative practices like sketching or prototyping support pathways thinking~\cite{Buxton-2014-SketchingUserDesign-d}, while iterative feedback and testing embody agentic thinking. Moreover, design thinking helps us \textit{reframe problems and overcome cognitive fixations}~\cite{jansson1991design, Lu-2017-SwitchingOnFixation-y,zamfirescu2021fake} about what goals are worth pursuing, which resonates with the emphasis on goal-setting in hope cognition. 

Thus, we ask: How hopeful is our research practice, really? Are we proactively exploring beyond surface-level or reactionary fixes? Do we feel empowered to define and take concrete next steps toward making meaningful changes? Do we engage in research practices that support adaptive strategies, reflection, and resilience over time? Specifically:
\begin{enumerate}
    \item Are we cultivating research environments that foster meaningful goal-setting, iterative reflection, and shared ownership over process and outcomes?
    \begin{itemize}
        \item Do our research goals strike the right balance between ambition and feasibility?
        \item Are we equipped with the tools to flexibly adjust our methods in response to obstacles or new insight?
        \item What are some challenges---such as limited student time, scarce resources, or prevailing publication norms---that might make conducting hopeful research difficult?
    \end{itemize}
    \item  Are we working in ways that center and sustain hope within the communities and participants we engage with?
    \begin{itemize}
        \item Are users and stakeholders truly co-creators in the process, empowered to shape both the goals and the pathways?
        \item What would it look like to measure or actively nurture such hope across research lifecycles?
    \end{itemize}
\end{enumerate}

By embracing a design mindset, one that values iteration, questioning assumptions, reimagining constraints, and embracing uncertainty, researchers can approach their work with a renewed sense of curiosity, adaptability, resilience, and hope. Hope here is not about believing in easy solutions or turning away from the complexity or challenges of the world. It is a practice of sitting with complexity, rethinking constraints, and holding onto the belief that new paths are still possible, whether through the tools and methods of research or through the solidarity built among researchers and communities working toward change~\cite{Hes-2014-DesigningHopeSustainability-m, McGeer-2004-ArtGoodHope-a, Bloch-1986-PrincipleHope-v}. Even if they do not solve everything or anything, they push researchers toward uncovering and taking the next best steps and actions that restore a sense of agency and make a better, even if uncertain, future feel possible~\cite{rylander2009design, cross2006designerly, Snyder-2002-HopeTheory-i}.

\subsection{Relevance to the CSCW Community}
We envision this workshop as a continuation of the \href{https://positech-cscw-2024.github.io/}{first Positech workshop} at CSCW 2024~\cite{positech}, with a focus on the methodological dimensions of hope-oriented design research. Many in the CSCW community are already engaged in tackling wicked~\cite{Zimmerman2007-ur} societal problems that resist simple answers and require interdisciplinary, reflective, and participatory approaches. This workshop aims to deepen that conversation by offering concrete practices and shared space for exploring how design methodologies can sustain hope, not simply as an emotional state but as a rigorous and deliberate way of making progress toward better futures in the face of complex societal problems. By drawing together researchers committed to social impact, we hope to foster new collaborations and frameworks for design that support collective resilience, agency, and the reimagining of what is possible.
\section{Organizers}
\textbf{JaeWon Kim} is a PhD candidate at the University of Washington Information School. Her research focuses on understanding, designing, and building social technologies that center on meaningful social connections, especially for the youth.

\textbf{Jiaying "Lizzy" Liu} is a PhD candidate at the School of Information at the University of Texas at Austin. Her research focuses on how multimodal technologies and video-sharing platforms shape online health discourse and care-seeking behaviors through a socio-technical lens.

\textbf{Cassidy Pyle} is a PhD candidate at the University of Michigan School of Information. Her research qualitatively explores how interactions with emerging technologies shape minoritized youth's access to college and career opportunities and well-being resources.

\textbf{Sowmya Somanath} is an Associate Professor at the University of Victoria. Her research focuses on understanding, designing, and building technologies that foster creativity and happiness in digital and tangible experiences.

\textbf{Lindsay Popowski} is a PhD candidate at Stanford University Computer Science. Her research is in the field of social computing systems. She works to design and build online spaces that facilitate relationship- and community-building, often by translating offline insights into new paradigms for the online world.

\textbf{Hua Shen} is a postdoctoral scholar at the University of Washington. Her research centers on bidirectional human-AI alignment, covering HCI and various AI fields. She works to empower humans to interactively explain, evaluate, collaborate with AI, and further provide human feedback to enhance AI development.
% She served as Associate Chairs for CHI, CHI LBW, Program Committees for ACL, EMNLP, and more.
% She received multiple awards, including AIED'24 Best Paper, CSCW'23 Best Demo, IUI'23 Best Paper Honorable Mention, 2023 Google Research Science Conference Scholarships, and 2023 Rising Stars of Data Science. 

\textbf{Casey Fiesler} is an Associate Professor of Information Science at University of Colorado Boulder. Her work focuses on technology ethics (including ethical speculation), as well as the good and bad of online communities.

\textbf{Gillian R. Hayes} is a Chancellor's professor and the Robert A. and Barbara L. Kleist Professor of Informatics in the Donald Bren School of Information and Computer Sciences at UC Irvine. Her work focuses on designing, developing, and testing supportive educational and health technologies with people who are traditionally left out of the design process.

\textbf{Alexis Hiniker} is an Associate Professor at the University of Washington Information School. She studies the way attention-economy design exploit users of all ages---but particularly children, teens, and families---and she designs more respectful alternatives to help people thrive. 

\textbf{Wendy Ju} is an Associate Professor in Information Science and Design Tech, and faculty in the Jacobs Technion-Cornell Institute at Cornell Tech. Her research focuses on designing interaction with automation, to understand the effects of context, culture and norms on interaction, to expand our capability to account for how interactions evolve over time, and leverage embedded computing to expand our ability to do interaction research at scale.

\textbf{Prof. Florian `Floyd' Mueller} is directing the Exertion Games Lab at Monash University in Melbourne, Australia, researching how interactive technologies can support human values. Floyd was general co-chair for CHI PLAY'18 and was selected to co-chair CHI'20 and CHI'24.

\textbf{Ahmer Arif} is an assistant professor at UT Austin's School of Information. His research focuses on understanding the spread of misinformation and designing based responses to that spread.

\textbf{Yasmine Kotturi} is an Assistant Professor of Human-Centered Computing at the University of Maryland, Baltimore County (UMBC), where she researches human-centered AI, design pedagogy, and community-driven methods for equitable futures of work.

% \input{sections/3_agenda}
% \section{Logistics}

% \subsection{Website}
% The workshop information is available at \href{website link}{https://www.google.com/}, where we provide a comprehensive overview of the workshop, including a call for participation, workshop agenda, and organizer information. Following the workshop, we intend to share its outcomes, such as the design method taxonomy and workshop submissions, upon author consent.

\section{Workshop Activities}
Participants will engage in four main activities: reframing problems to uncover new possibilities for action; analyzing design methods to identify what works, what is missing, and how methods can better support hope; discussing how knowledge-sharing and community-building can extend the impact of design practices; and personally reflecting on how to apply these ideas in their own research and practice. The workshop closes with a collective discussion on key takeaways and ways to carry this work forward.

\section{Post-Workshop Plan}
Looking beyond the workshop, our goal is to grow and sustain the Positech community as a space for ongoing dialogue and collaboration among researchers working in similar areas and goals. We also plan to develop a comprehensive taxonomy of generative and iterative design research methods and share it on our website.

\begin{acks}
JaeWon Kim would like to acknowledge the CERES Network, University of Washington Global Innovation Funds (GIF) and Student Technology Funds (STF), which provided support for this work. Florian `Floyd' Mueller thanks the Australian Research Council, especially DP190102068, DP200102612 and LP210200656. Alexis Hiniker is a special government employee for the Federal Trade Commission. The content expressed in this manuscript does not reflect the views of the Commission or any of the Commissioners.
\end{acks}

\bibliographystyle{ACM-Reference-Format}
\bibliography{references}

\end{document}